\documentclass[aps,preprint,prd,showpacs,nofootinbib]{revtex4-1}
\usepackage{graphicx}
\usepackage{bm}
\usepackage{times}
\usepackage[hypertex]{hyperref}
\usepackage{slashed}
\def\br{\begin{eqnarray}}
\def\er{\end{eqnarray}}
\def\be{\begin{equation}}
\def\ee{\end{equation}}

\def\({\left(}
\def\){\right)}

\begin{document}

\title{WIMPs in a 3-3-1 model with heavy sterile neutrinos}

\author{J. K. Mizukoshi$^a$}\email{mizuka@ufabc.edu.br}  \author{C. A. de S. Pires$^{b}$}\email{cpires@fisica.ufpb.br} \author{F. S. Queiroz$^{b}$}\email{farinaldo@fisica.ufpb.br} \author{P. S. Rodrigues da Silva$^{b}$}\email{psilva@fisica.ufpb.br}
\affiliation{$^a$Centro de Ci\^encias Naturais e Humanas, Universidade Federal
do ABC,
R. Santa Ad\'elia 166, 09210-170, Santo Andr\'e - SP, Brazil.
\\
$^{b}$Departamento de F\'{\i}sica, Universidade Federal da
Para\'{\i}ba, 
Caixa Postal 5008, 58051-970, Jo\~ao Pessoa - PB, Brazil.}

\date{\today}
\vspace{1cm}
\begin{abstract}
In this work we build a gauge model based on the $SU(3)_c\otimes SU(3)_L\otimes
U(1)_N$ symmetry with heavy neutrinos and show that we can have two weakly interacting cold dark matter candidates in its spectrum. This is achieved by noticing that a global $U(1)$ symmetry can be imposed on the model in such a way that the stability of the dark matter is guaranteed.
We obtain their relic abundance and analyze their compatibility with recent direct detection experiments, also exploring the possibility of explaining the two events reported by CDMS-II. An interesting outcome of this 3-3-1 model, concerning direct detection of these WIMPs, is a strong bound on the symmetry breaking scale, which imposes it to be above 3~TeV.
\end{abstract}
    
\maketitle
\section{Introduction}

The Dark Matter (DM) problem constitutes a key problem at the interface among particle physics, astrophysics and cosmology. The observational data accumulated in the last decade point to the existence of a non-negotiable amount of non-baryonic DM, whose identity is still unknown. Since the standard model of electroweak interactions (SM) does not provide any candidate for such invisible component of matter, this problem is an indication for physics beyond the SM. Different measurements coming from cosmic microwave background radiation (CMBR)~\cite{wmap}, galaxy rotation curves~\cite{rota}, gravitational lensing~\cite{lentes} and structure formation~\cite{formation} etc., confirm that, besides its undoubted evidence, it must contribute to around $22\%$ of the total energy density of the Universe. Nowadays it is  known, due to numerical simulations to reproduce the structure formation~\cite{fulano1}, that the matter present in our Universe is dominated by cold dark matter (CDM), and precise measurements of its relic abundance impose strong constraints on various new physics models. However, the relic abundance alone is not enough to point out all the properties of the CDM, even already assuming that the CDM is represented by a weakly interacting massive particle (WIMP). Colliders, and particularly the Large Hadron Collider (LHC), have the potential for discovering and identifying new particles not predicted by the SM. By the other side, most of the information about the nature of the CDM might be extracted from existing direct detection experiments, through the scattering of WIMPs with nuclei~\cite{fulano2} and therefore we will focus on them. 

The WIMPs are the most studied CDM candidates and arise naturally in several theoretical frameworks such as Supersymmetry~\cite{supersimetria1,supersimetria2}, Universal Extra Dimensions~\cite{dimensoesextras}, Little Higgs Models~\cite{littlehiggs}, Technicolor~\cite{TC}, etc., but since all these theories remain hypothetical~\cite{bottino}, it is equally worthwhile to tackle less conventional possibilities. For this reason, we are going to explore a small gauge extension of the electroweak sector of SM, $SU(3)_C\otimes SU(3)_L\otimes U(1)_N$, 3-3-1 for short (for a nice review see Ref.~\cite{331review}).
This extension can be accomplished by a class of models~\cite{models} that have intriguing features such as: the models are anomaly free only if the number of families is a multiple of three allied to the condition of QCD asymptotic freedom~\cite{numerodefamilias};  the electric charge quantization and the explanation of the vector-like nature of the electromagnetism are  naturally achieved in the absence of anomalies~\cite{eletromagntismo}; there is room for lepton number violation~\cite{leptonnumber} and new sources of CP violation~\cite{martinez}, crucial features to approach baryogenesis and/or leptogenesis, among other nice characteristics of the model.

The CDM problem in 3-3-1 models was already studied in different situations, with self-interacting DM~\cite{selfinteracting}, a scalar bilepton (a particle that carries two units of lepton number) WIMP~\cite{JCAP} and a supersymmetric self-interacting DM~\cite{longcdm}. Here we are going to consider a variation of the model of Ref.~\cite{JCAP} and make a deep analysis of the CDM candidates and their respective abundance and direct detection. In the work of Ref.~\cite{JCAP} there are light right-handed neutrinos in the triplet representation of $SU(3)_L$ and no singlet neutrino, while the version of the model developed here contains new left-handed neutrinos in the fundamental representation of $SU(3)_L$ instead of right-handed neutrinos. For this reason we will call it 3-3-1LHN for short. This 3-3-1LHN model was inspired by the first attempts to enlarge the electroweak gauge symmetry $SU_L(2)\otimes U_Y(1)$ to $SU_L(3)\otimes U_N(1)$~\cite{valle}.  It is amazing that a global $U(1)$ symmetry can be imposed that not only simplifies the Yukawa Lagrangian and the scalar potential but also stabilizes the lightest of the new particles charged under this symmetry, providing candidates for explaining the CDM problem.
Our main goal is to get WIMP CDM candidates from the spectrum of 3-3-1LHN in agreement with most recent bounds from  direct detection experiments, namely CDMS-II~\cite{cdms} and XENON~\cite{xenon}, and investigate the region of parameter space which is well suited for explaining the positive signals observed by the CDMS-II experiment~\cite{kopp}. 
We do not take into account DAMA~\cite{DAMA} and CoGeNT~\cite{CoGeNT} results in this work since our model does not have any allowed region in the parameter space that explains the observed signals by these experiments. Nevertheless, we remark that the parameter space favored by these experiments is mostly in conflict with all other detection experiments, although it remains an intriguing challenge to be solved.

We start by briefly describing the model in section~\ref{sec1}, introducing its main ingredients. In section~\ref{sec3} we discuss the relic abundance computation as well as the direct detection bounds for the CDM candidates and analyze the compatibility of our model with the positive signal from CDMS-II. We present our conclusions in section~\ref{sec4}.

\section{The 3-3-1LHN Model}
\label{sec1}

In the 3-3-1LHN model the leptons are accommodated  in triplet and singlet representations as follows (we indicate the $SU(3)_C\otimes SU(3)_L\otimes U(1)_N$ transformation properties in parentheses),

\begin{eqnarray}
f_{aL} = \left (
\begin{array}{c}
\nu_{a} \\
e_{a} \\
N_{a}
\end{array}
\right )_L\sim(1\,,\,3\,,\,-1/3)\,,\,\,\,e_{aR}\,\sim(1,1,-1)\,,\,\,\,N_{aR}\,\sim(1,1,0),
\label{L}
\end{eqnarray}
where $a=1,2,3$ represents the family index for the usual three generation of leptons, while $N_{a(L,R)}$ are new heavy neutrinos representing new degrees of freedom in this model, and it is this assumption that makes the 3-3-1LHN model substantially different from the proposal studied before~\cite{JCAP}.

In the Hadronic sector, the first generation comes in the triplet representation and the other two are in an anti-triplet representation of $SU_L(3)$, as a requirement for anomaly cancellation. They are given by,
\begin{eqnarray}
&&Q_{iL} = \left (
\begin{array}{c}
d_{i} \\
-u_{i} \\
d^{\prime}_{i}
\end{array}
\right )_L\sim(3\,,\,\bar{3}\,,\,0)\,, \nonumber \\
&&
u_{iR}\,\sim(3,1,2/3),\,\,\,
\,\,d_{iR}\,\sim(3,1,-1/3)\,,\,\,\,\, d^{\prime}_{iR}\,\sim(3,1,-1/3),\nonumber \\
&&Q_{3L} = \left (
\begin{array}{c}
u_{3} \\
d_{3} \\
u^{\prime}_{3}
\end{array}
\right )_L\sim(3\,,\,3\,,\,1/3)\,, \nonumber \\
&&
u_{3R}\,\sim(3,1,2/3),
\,\,d_{3R}\,\sim(3,1,-1/3)\,,\,u^{\prime}_{3R}\,\sim(3,1,2/3)
\label{quarks} 
\end{eqnarray}
where the index $i=1,2$ where chosen to represent the first two generations. The primed quarks are new heavy quarks with the usual fractional electric charges.

In order to generate SM fermion masses, three scalar triplets are introduced, 
\begin{eqnarray}
\eta = \left (
\begin{array}{c}
\eta^0 \\
\eta^- \\
\eta^{\prime 0}
\end{array}
\right ),\,\rho = \left (
\begin{array}{c}
\rho^+ \\
\rho^0 \\
\rho^{\prime +}
\end{array}
\right ),\,
\chi = \left (
\begin{array}{c}
\chi^0 \\
\chi^{-} \\
\chi^{\prime 0}
\end{array}
\right )\,, 
\label{scalarcont} 
\end{eqnarray}
with $\eta$ and $\chi$ both transforming as $(1\,,\,3\,,\,-1/3)$
and $\rho$ transforming as $(1\,,\,3\,,\,2/3)$.

In general, a discrete $Z_2$ symmetry is usually assumed, transforming the fields as,
\begin{eqnarray}
\left( \chi\,,\,\rho\,,e_{aR}\,,\,N_{aR}\,,\, u_{aR}\,,\,d^{\prime}_{iR}\,,\, Q_{3L} \right) \rightarrow -\left( \chi\,,\,\rho\,,e_{aR}\,,\,N_{aR}\,,\, u_{aR}\,,\,d^{\prime}_{iR}\,,\, Q_{3L}\right)\,, 
	\label{discretesymmetryI}
\end{eqnarray}
which leads to an economical model with a simplified Yukawa Lagrangian~\footnote{ We are going to see that this discrete symmetry can be replaced by a global $U(1)$ symmetry, with the effect of producing the same terms in the Lagrangian of the model, but also advantageous in stabilizing our CDM candidates.},
\begin{eqnarray}
&-&{\cal L}^Y =f_{ij} \bar Q_{iL}\chi^* d^{\prime}_{jR} +f_{33} \bar Q_{3L}\chi u^{\prime}_{3R} + g_{ia}\bar Q_{iL}\eta^* d_{aR} \nonumber \\
&&+h_{3a} \bar Q_{3L}\eta u_{aR} +g_{3a}\bar Q_{3L}\rho d_{aR}+h_{ia}\bar Q_{iL}\rho^* u_{aR} \nonumber \\
&&+ G_{ab}\bar f_{aL} \rho e_{bR}+g^{\prime}_{ab}\bar{f}_{aL}\chi N_{bR}+ \mbox{H.c}. 
\label{yukawa}
\end{eqnarray}
Again, in these expressions we are using the family indices $i=1,2$ and $a=1,2,3$.

The most general scalar potential that we can construct which obeys the above discrete symmetry has the form,
\begin{eqnarray} V(\eta,\rho,\chi)&=&\mu_\chi^2 \chi^2 +\mu_\eta^2\eta^2
+\mu_\rho^2\rho^2+\lambda_1\chi^4 +\lambda_2\eta^4
+\lambda_3\rho^4+ \nonumber \\
&&\lambda_4(\chi^{\dagger}\chi)(\eta^{\dagger}\eta)
+\lambda_5(\chi^{\dagger}\chi)(\rho^{\dagger}\rho)+\lambda_6
(\eta^{\dagger}\eta)(\rho^{\dagger}\rho)+ \nonumber \\
&&\lambda_7(\chi^{\dagger}\eta)(\eta^{\dagger}\chi)
+\lambda_8(\chi^{\dagger}\rho)(\rho^{\dagger}\chi)+\lambda_9
(\eta^{\dagger}\rho)(\rho^{\dagger}\eta) \nonumber \\
&&-\frac{f}{\sqrt{2}}\epsilon^{ijk}\eta_i \rho_j \chi_k +\mbox{H.c}.
\label{potential}
\end{eqnarray}
It is well known that this potential is appropriate to induce the desired spontaneous symmetry breaking pattern of the electroweak gauge symmetry, $ SU(3)_L\otimes U(1)_N$ to $SU(2)_L\otimes U(1)_Y$ and finally to $U(1)_{QED}$, generating the masses of gauge bosons and fermions through the Higgs mechanism.

We also write the currents involving the non-hermitian vector bosons for leptons and quarks, since it is going to be necessary in observing the existence of an extra global symmetry in the 3-3-1LHN model. It reads (see the fourth paper in Ref.~\cite{models}),
\br
{\cal L}_{NH} &=&
-\frac{g}{\sqrt{2}}\left[\bar{\nu}^a_L\gamma^\mu e_L^a W^+_\mu +\bar{N}_L^a\gamma^\mu e_L^a V^+_\mu + \bar{\nu}^a_L\gamma^\mu N_L^a U^0_\mu  \right. \nonumber \\
&& \left. +\left(\bar{u}_{3L}\gamma^\mu d_{3L}  +\bar{u}_{iL}\gamma^\mu d_{iL}\right)W_\mu^+ +\left(\bar{u}^\prime_{3L}\gamma^\mu d_{3L}  +\bar{u}_{iL}\gamma^\mu d^\prime_{iL}\right)V_\mu^+  \right. \nonumber \\
&& \left. +\left(\bar{u}_{3L}\gamma^\mu u^\prime_{3L}  -\bar{d}^\prime_{iL}\gamma^\mu d_{iL}\right)U_\mu^0
+ {\mbox h.c.}
\right]\,,
\label{CC} 
\er
where we have defined $W_\mu^+ = \frac{1}{\sqrt{2}}(W_\mu^1-iW_\mu^2)$, as usual, $V_\mu^-=\frac{1}{\sqrt{2}}(W_\mu^6-iW_\mu^7)$ and $U_\mu^0=\frac{1}{\sqrt{2}}(W_\mu^4-iW_\mu^5)$. The three remaining neutral gauge bosons, $A_\mu$, $Z_\mu$ and $Z^\prime_\mu$, couple to the fermions in a diagonal basis and do not influence the discussion on the new symmetry that follows, so we do not present their currents here.

Now we notice that there exists a new extra global symmetry in this model, which we call $U(1)_G$, with the following assignments of $\mathbf{G}$ charges carried exclusively by the 3-3-1 model fields,
\be
\mathbf{G}({\bar N}_{L/R},\,{\bar u}_{3L/R}^\prime,\,d_{iL/R}^\prime,\,V_\mu^-,\,U_\mu^{0},\,\chi^{0},\,\chi^{-},\,\eta^{\prime 0 *},\,\rho^{\prime -})=+1\,.
\label{LNA}
\ee
All the other fields transform trivially under this symmetry.
At this point we notice that we could have started with this $U(1)_G$ global symmetry from the beginning, without imposing the previously mentioned discrete symmetry, Eq.~(\ref{discretesymmetryI}). In other words, if we replace the $Z_2$ global symmetry by the $U(1)_G$ global symmetry, we recover the same Lagrangian terms as given in Eq.~(\ref{yukawa}) and Eq.~(\ref{potential}) with no new term to be added. The advantage of this continuous symmetry is that the $\mathbf{G}$ charged fields (we call them $\mathbf{G}$-fields for short) always appear in pairs, guaranteeing that the lightest one is stable. Next we identify the mass eigenstates in the 3-3-1LHN model such as to select which neutral $\mathbf{G}$-fields can be a potential CDM candidate~\footnote{We present the trilinear couplings for the $\mathbf{G}$-fields in appendix {\bf A}.}.

\subsection{The mass eigenstates}

In order to achieve spontaneous symmetry breaking, we suppose that the neutral scalars ($\eta^0 ,\, \rho^0 ,\, \chi^{\prime 0})$ develop a vacuum expectation value (VEV) according to,
\begin{eqnarray}
 \eta^0 , \rho^0 , \chi^{\prime 0} \rightarrow  \frac{1}{\sqrt{2}} (v_{\eta ,\rho ,\chi^{\prime}} 
+R_{ \eta ,\rho ,\chi^{\prime}} +iI_{\eta ,\rho ,\chi^{\prime}})\,, 
\label{vacua} 
\end{eqnarray}
where we make the reasonable and simplifying  assumption that the remaining neutral scalars $(\eta^{\prime 0}, \chi^{0})$ do not develop VEVs~\footnote{If we take non-trivial VEVs for these scalars we would still obtain the complete mass spectrum of the model with only additional complexity in the mixing of gauge bosons and scalars. However, this would also break the $U(1)_G$ global symmetry, yielding an unwanted Goldstone boson in the spectrum.}.

From this pattern of symmetry breaking, we observe that   
the $U(1)_G$ symmetry forbids Majorana mass terms for the neutrinos and no mixing appears among the new neutrinos with the standard ones. This turns them into truly sterile Dirac neutrinos. Moreover, for sake of simplicity, we consider that the mass matrix of the charged leptons, new neutrinos  and of the new quarks all come in diagonal mass bases with normal hierarchy.

Considering the vacuum structure in Eq.~(\ref{vacua}), the  mass matrix of the new neutrinos and quarks take the form,
\be
M_{Na}=\frac{g^{\prime}_{aa}}{\sqrt{2}}v_{\chi^{\prime}}\,,
\label{mneut}
\ee
and  
\be
M_{q^{\prime}_a}= \frac{f_{aa}}{\sqrt{2}}v_{\chi^{\prime}}\,,
\label{mq}
\ee
respectively. If we assume that $g^\prime_{11} < g^\prime_{22} \leq g^\prime_{33}$ in the first of these equations, the lightest heavy neutrino is identified with $N_1$. 
Regarding the standard neutrinos, we  assume here that their tiny masses are due to effective dimension-5 operators as first implemented in Ref.~\cite{lightnu}.

As for the scalar mass matrices, we first need the minimum conditions from the potential in Eq.~(\ref{potential}), given by, 
\begin{eqnarray}
 &&\mu^2_\chi +\lambda_1 v^2_{\chi^{\prime}} +
\frac{\lambda_4}{2}v^2_\eta  +
\frac{\lambda_5}{2}v^2_\rho-\frac{f}{2}\frac{v_\eta v_\rho}
{ v_{\chi^{\prime}}}=0,\nonumber \\
&&\mu^2_\eta +\lambda_2 v^2_\eta +
\frac{\lambda_4}{2} v^2_{\chi^{\prime}}
 +\frac{\lambda_6}{2}v^2_\rho -\frac{f}{2}\frac{v_{\chi^{\prime}} v_\rho}
{ v_\eta} =0,
\nonumber \\
&&
\mu^2_\rho +\lambda_3 v^2_\rho + \frac{\lambda_5}{2}
v^2_{\chi^{\prime}} +\frac{\lambda_6}{2}
v^2_\eta-\frac{f}{2}\frac{v_\eta v_{\chi^{\prime}}}{v_\rho} =0.
\label{mincond} 
\end{eqnarray}

Although the trilinear coupling $f$ in Eq.~(\ref{potential}) is a free mass parameter, in this work we make the assumption that $f$ is of the order of the 3-3-1 symmetry breaking scale, $v_{\chi^{\prime}}$, supposed to be at TeV scale,
while $v_\rho$ and $v_\eta$ ($\ll v_{\chi^{\prime}}$) have to be at the electroweak breaking scale, $v\approx 246$~GeV, since they fix the $Z$ and $W^\pm$ gauge boson masses~\cite{models}, being related by $v^{2}_{\eta} + v^{2}_\rho=v^{2}$. We then choose $f=\frac{v_{\chi^{\prime}}}{2}$ and $v_\rho = v_\eta = \frac{v}{\sqrt{2}}$, just to simplify the diagonalization procedure of the scalar mass matrices.

Substituting the Eqs.~(\ref{vacua}) and (\ref{mincond}) into the scalar potential, Eq.~(\ref{potential}), we can obtain the mass matrices for the neutral scalars in three different bases, a scalar, $(R_{\chi^{\prime}}\,,\,R_\eta\,,\,R_\rho)$, a pseudo-scalar one, $(I_{\chi^{\prime}}\,,\,I_\eta\,,\,I_\rho)$, and a complex scalar basis, $( \chi^{0\dagger}\,,\,\eta^{\prime 0})$. 

Since no fine-tuning is assumed we can take some simplifying relations here in order to obtain the mass eigenstates. Namely, $\lambda_{4}=\lambda_{5}=0.25$ and  $\lambda_{2}=\lambda_{3}$. This assumption, which in principle would demand some kind of symmetry to guarantee such equalities, may have some implication on the mixing of interaction eigenstates which could change somehow our results. However, at this point we still do not have a consistent way of considering more general scenarios where this diagonalization can be numerically implemented, an issue we hope to develop in the future. For this reason we are going to perform our computations in this framework, keeping in mind that some different outcome could emerge in a more general scheme, which would have our scenario as a subset of possibilities.

We then find the following mass eigenvectors in the basis $(R_{\chi^{\prime}}\,,\,R_\eta\,,\,R_\rho)$,
\begin{eqnarray}
S_1   & = & R_{\chi^{\prime}}\,,\nonumber \\ 
 S_2 & = &  \frac{1}{\sqrt{2}}(R_\eta-R_\rho)\,,\\ \nonumber
 H  & = &  \frac{1}{\sqrt{2}}(R_\eta+R_\rho),
\label{eigenvextorsRe}
\end{eqnarray}
with respective mass eigenvalues~\footnote{We notice that the real and pseudo-scalar mass eigenvalues in Ref.~\cite{JCAP} are lacking a factor of two, while the WIMP complex scalar has a correct factor. This does not change the qualitative results and conclusions in that work, although tiny quantitative corrections are implied wherever the Higgs boson play some role. Here we took those missing factors into account.},
\begin{eqnarray}
M^{2}_{S_{1}} & = & \frac{v^{2}}{4}+2v_{\chi^\prime}^{2}\lambda_{1}\,, \nonumber \\
M^{2}_{S_{2}} & = & \frac{1}{2}(v_{\chi^\prime}^{2}+2v^{2}(2\lambda_{2}-\lambda_{6}))\,, \nonumber \\
M^{2}_{H} & = & v^{2}(2\lambda_{2}+\lambda_{6})\,.
\label{massashiggs}
\end{eqnarray}

In the pseudo-scalar basis, $(I_{\chi^{\prime}}\,,\,I_\eta\,,\,I_\rho)$ we find the mass eigenstates,
\begin{eqnarray}
I^{0}_{1} & = & - \frac{ 1 }{ \sqrt{ 1+\frac{v^{2}}{v_{\chi^\prime}^{2}} } }\ I_{\chi^{\prime}} +\frac{ v }{ v_{\chi^\prime}\sqrt{ 1+\frac{v^{2}}{v_{\chi^\prime}^{2}} } }\ I_\rho \,,\nonumber \\
I^{0}_{2} & = & \frac{1}{\sqrt{2}}(-\frac{v_{\chi^\prime}}{v}+ \frac{ v_{\chi^\prime} }{ v( 1+\frac{v^{2}}{v_{\chi^\prime}^{2}} ) })\ I_{\chi^{\prime}} +  \frac{1}{\sqrt{2}}\ I_\eta - \frac{1}{\sqrt{2}( 1+\frac{v^{2}}{v_{\chi^\prime}^{2}} )}\ I_\rho \,,\nonumber \\
P_1 & = & \frac{ v }{ v_{\chi^\prime}\sqrt{ 2+\frac{v^{2}}{v_{\chi^\prime}^{2}} } }\ I_{\chi^{\prime}} + \frac{ 1 }{\sqrt{ 2+\frac{v^{2}}{v_{\chi^\prime}^{2}} } }\ I_{\eta}+\frac{ 1 }{\sqrt{ 2+\frac{v^{2}}{v_{\chi^\prime}^{2}} } }\ I_{\rho},
\end{eqnarray}
where, $I^{0}_{1}$ and $I^{0}_{2}$, correspond to Goldstone bosons and $P_1$ is a massive pseudo-scalar that remains in the spectrum whose mass is,
\begin{eqnarray}
M^{2}_{P_{1}} = \frac{1}{2}(v_{\chi^\prime}^{2}+\frac{v^{2}}{2}).
\end{eqnarray}

Also, in the basis of complex neutral scalars, $( \chi^{0}\,,\,\eta^{\prime 0 *})$, we get the mass eigenstates,
\begin{eqnarray}
G_\phi    & = &  -\frac{v_{\chi^\prime}}{v\sqrt{1+\frac{v_{\chi^\prime}^2}{v^2}}}\ \chi^{0} + \frac{1}{\sqrt{1+\frac{v_{\chi^\prime}^2}{v^2}}}\ \eta^{\prime 0 *} \,,\nonumber \\
\phi & = & \frac{v}{ v_{\chi^\prime}\sqrt{1+\frac{v^2}{v_{\chi^\prime}^2}}}\ \chi^{0*} +\frac{1}{\sqrt{1+\frac{v^2}{v_{\chi^\prime}^2}}}\ \eta^{\prime 0},
\label{wimp}
\end{eqnarray}
where $G_\phi$ is recognized as the Goldstone boson eaten by the gauge bosons $U^{0}$ and $U^{0 \star}$ and $\phi$ has a mass,
\begin{eqnarray}
M^{2}_{\phi} & = & \frac{(\lambda_{7} + \frac{1}{2} )}{2}[v^{2}+v_{\chi^\prime}^{2}].
\end{eqnarray}

Considering the two bases of charged scalars, $(\chi^{-}\,,\,\rho^{\prime -})$ and $(\eta^{-} \,,\, \rho^{-})$, we obtain the following mass eigenstates,
\begin{eqnarray}
h^-_1 & = & \frac{1}{\sqrt{1+\frac{v^2}{v_{\chi^\prime}^2}}}(\frac{v}{v_{\chi^\prime}}\chi^{-} + \rho^{\prime-})\,, \nonumber \\
h^-_2 & = & \frac{1}{\sqrt{2}}(\eta^{-} + \rho^{-} )\,,
\label{vetoresCS}
\end{eqnarray} 
which can be checked to be the same eigenvectors as in Ref.~\cite{JCAP} when we take the limit $v_{\chi^\prime} \gg v$. Their mass eigenvalues are,
\begin{eqnarray}
M^{2}_{h^{-}_{1}} & = & \frac{\lambda_{8}+\frac{1}{2} }{2}(v^{2}+v_{\chi^\prime}^{2})\,, \nonumber \\
M^{2}_{h^{-}_{2}} & = & \frac{v_{\chi^\prime}^{2}}{2}+\lambda_{9}v^{2}\,.
\label{massash1h2}
\end{eqnarray}
The remaining eigenvectors are two Goldstone bosons given by,
\begin{eqnarray}
h^-_3 & = & \frac{1}{\sqrt{1+\frac{v^2}{v_{\chi^\prime}^2}}}(\chi^{-} -\frac{v}{v_{\chi^\prime}}\rho^{\prime-})\,, \nonumber \\
h^-_4 & = & \frac{1}{\sqrt{2}}(\eta^{-} - \rho^{-} ).
\label{vetoresCS2}
\end{eqnarray}

Finally, from the gauge invariant scalar kinetic terms (not shown here) and using Eq.~(\ref{vacua}), we easily obtain the gauge boson masses~\cite{models},
\begin{eqnarray}
m_{W^\pm}^2    &=& \frac{1}{4}g^2v^2\,,
\nonumber \\
m^{2}_{Z}      &=& m_{W^\pm}^2/c^{2}_{W}\,,
\nonumber \\
m^2_{V^\pm}    &=& m^2_{U^0} = \frac{1}{4}g^2(v_{\chi^\prime}^2+v^2)\,,
\nonumber \\
m^2_{Z^\prime} &=& \frac{g^{2}}{4(3-4s_W^2)}[4c^{2}_{W}v_{\chi^\prime}^2 +\frac{v^{2}}{c^{2}_{W}}+\frac{v^{2}(1-2s^{2}_{W})^2}{c^{2}_{W}}]\,,
\label{massvec}
\end{eqnarray}
where we have defined the Weinberg mixing angle through $\sin\theta_W\equiv s_W$ (as well as $\cos\theta_W\equiv c_W$). Notice that we have neglected the mixing between the neutral gauge bosons $Z$ and $Z^\prime$, which is constrained to be very small (see the fourth paper in Ref.~\cite{models}).
 
With all the mass eigenstates identified as above we are able to consider the stability of the neutral $\mathbf{G}$-fields. In the 3-3-1 model with right-handed neutrinos studied in Ref.~\cite{JCAP}, $\phi$ was the same combination of interacting neutral scalar $\mathbf{G}$-fields as in this 3-3-1LHN model, but there this scalar carried two units of lepton number instead. There is no other neutral $\mathbf{G}$-field scalar in the 3-3-1LHN model, and the only neutral $\mathbf{G}$-field vector boson is $U^0$. These, together with the lightest heavy $\mathbf{G}$-field neutrino, $N_1$, are the potential CDM candidates of this model, although they cannot be simultaneous candidates since they couple to each other plus some standard model particle, as explicitly shown for the trilinear couplings in appendix {\bf A}. Thus, it is enough to make one of them the lightest particle among the $\mathbf{G}$-fields, which provides a stable CDM candidate.
We would like to stress that although the gauge boson, $U^0$, could be the stable $\mathbf{G}$-field, it leads to a too much suppressed relic abundance, thus we do not consider it henceforth as a third CDM candidate.

Having found the mass spectrum of the 3-3-1LHN model we identify $\phi$ and $N_1$ as our possible DM candidates (again, $U^0$ is a candidate too, but extremely underabundant) by enforcing that one of them be the lightest $\mathbf{G}$-field, we will next determine their relic abundance and analyze the WMAP favored parameter space region under direct detection experiments.

\section{Relic Abundance and Direct Detection} 
\label{sec3}

Among the CDM candidates, the WIMPs are the most intriguing ones since their thermal cross section, which is roughly at the electroweak scale, naturally leads to the appropriate relic density. The scenario goes as follows: a WIMP which is in thermodynamic equilibrium with the plasma in the early Universe decouples when its interaction rate drops below the expansion rate of the Universe. 
In this way we have first to check that the CDM candidate besides being stable (or meta-stable), either freezes with the right relic abundance~\cite{wmap} or, at least,  represents the majority of CDM constituting a subdominant scenario. Secondly, since nowadays we have some direct detection experiments available~\cite{cdms,xenon}, it would be desirable that our candidate has at least some chance of being detected in the near future or, more remarkably, to explain positive signals such as the events in excess observed by CDMSII. 

First we will describe the computational procedure used to get the relic abundance of the 3-3-1LHN CDM candidates, $\phi$ and $N_{1}$, and present some scatter plots showing our results for different regimes of the parameter space. Lastly, we will discuss a little bit about the direct detection method and compute the WIMP-nucleon cross section of our candidates and investigate its feasibility in light of CDMS and XENON bounds and also the possibility of explaining the recent CDMSII signal.

\subsection{Relic Abundance}

In order to obtain the WIMP abundance in its decoupling stage we need to solve the Boltzmann equation which gives the evolution of the abundance of a generic species in the Universe as a function of the temperature,
\begin{equation}
\frac{dY}{dT}=\sqrt{\frac{\pi g_{\ast}(T)}{45}}M_{p} -\left\langle \sigma v\right\rangle(Y^{2}-Y^{2}_{eq})
\end{equation}
where, $g_{\ast}$ is the effective number of degrees of freedom available at the freeze-out temperature, $M_{p}$ is the Planck mass, $Y$ is the thermal abundance or number density over entropy  (while $Y_{eq}$ is the abundance at the equilibrium epoch) and $\left\langle \sigma v\right\rangle$ is the thermal averaged cross section for WIMP annihilation times the relative velocity. The particle physics information of the model enters in this cross-section which includes all annihilation and co-annihilation channels,
\begin{equation}       
\left\langle \sigma v\right\rangle=\frac{\displaystyle\sum_{i,j}g_{i}g_{j}\displaystyle\int_{ (m_{i}+m_{j})^{2} }ds\sqrt{s}K_{1}(\frac{\sqrt{s}}{T})p_{ij}^{2}\displaystyle\sum_{k,l}\sigma_{ij;kl}(s)}{2T(\displaystyle\sum_{i}g_{i}m^{2}_{i}K_{2}(m_{i}/T) )^{2}},
\end{equation}
where $g_{i}$ is the number of degrees of freedom that characterizes the species involved, $\sigma_{ij;kl}$ the total cross-section for annihilation of a pair of particles with masses $m_{i}$, $m_{j}$ into some SM particles $(k,l)$ with respective masses $m_{k}$ and $m_{l}$, while $p_{ij}$ is the momentum of incoming particles in their center of momentum frame.

\begin{figure}[htb]
\centering
\includegraphics[width=0.6\columnwidth]{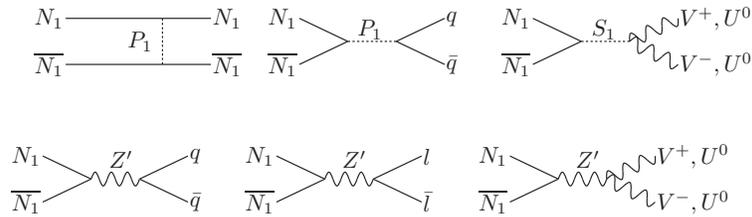}
\caption{The main processes which contribute to the abundance of $N_{1}$, where $l=e,\mu,\tau,n_{e},n_{\mu},n_{\tau}$ and $q=u,d,c,s,t,b$.}
\label{processoN1}
\end{figure}

\begin{figure}[htb]
\centering
\includegraphics[width=0.6\columnwidth]{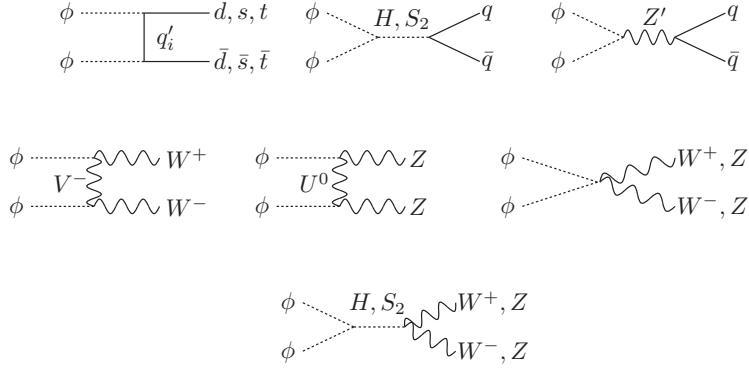}
\caption{The main processes which contribute to the abundance of $\phi$}
\label{processofi} 
\end{figure}

\begin{figure}[!htb]
\centering
\includegraphics[width=0.49\columnwidth]{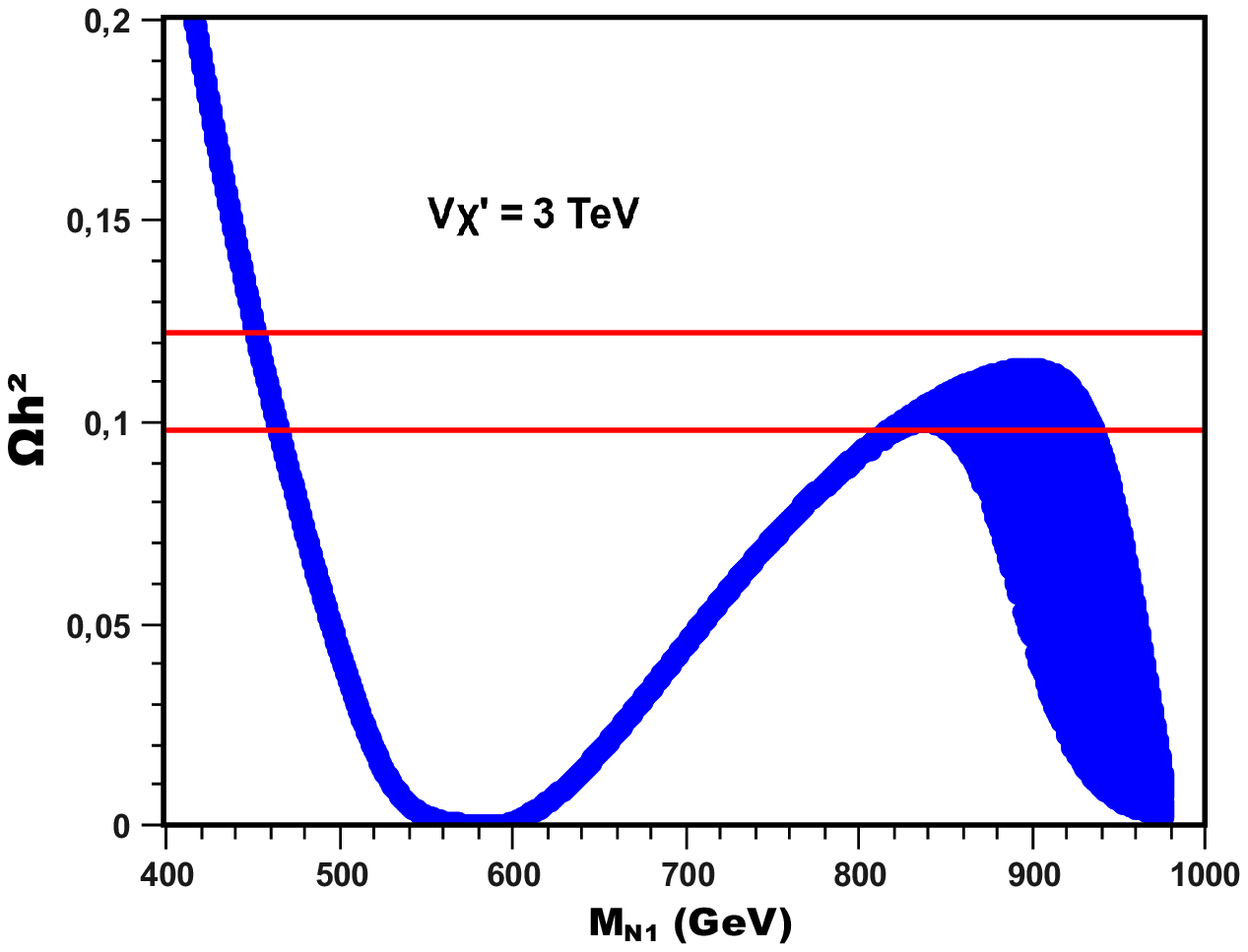}
\includegraphics[width=0.49\columnwidth]{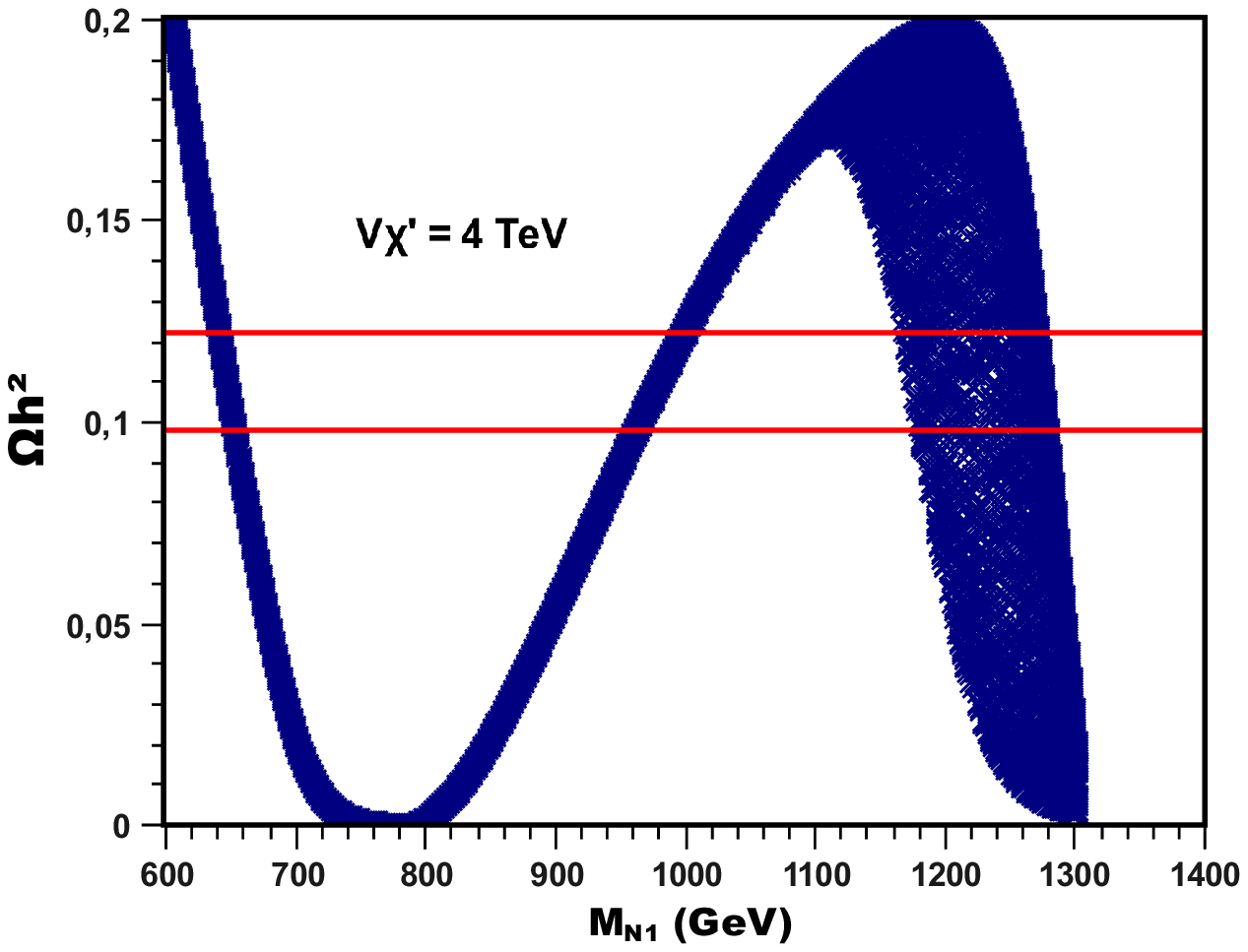}
\caption{Relic abundance for the heavy neutrino $N_{1}$ with the region in accordance with WMAP7, $0.098 \leq \Omega h^2 \leq 0.122$, shown between the red bars. We used $400\ \mbox{GeV} \leq M_{S1} \leq 4.5\ \mbox{TeV}$ and $v_{\chi^\prime} = 3\ \mbox{TeV}$ in the left panel and $600\ \mbox{GeV} \leq M_{S1} \leq 6\ \mbox{TeV}$ and $v_{\chi^\prime} = 4\ \mbox{TeV}$ in the right one.}
\label{abundanciaN1}
\end{figure}

The relic density is obtained by integrating from $T=\infty$ to $T=T_{0}$ where $T_{0}$ is the temperature of the Universe today, yielding,
\begin{equation} 
\Omega h^{2}=2.742\times 10^{8}\frac{M_{WIMP}}{\mbox{GeV}} Y(T_{0})
\end{equation}
Our results are obtained by using the package micrOMEGAs~\cite{micromegas}, which computes this relic density numerically for a given model. The task would reveal unfeasible analytically since many interactions participate in the annihilation process at freeze-out. We have also implemented the 3-3-1LHN model in the package lanHEP~\cite{lanhep} that furnishes the model files to be used in micrOMEGAs, making the task of computing the relic density much easier and reliable. The most significant processes which contribute to the abundance of our CDM candidates, $N_{1}$ and $\phi$, separately,  are shown in Figs.~\ref{processoN1}-\ref{processofi}.
After using the procedure described above we then show the results for each candidate. In Fig.~\ref{abundanciaN1} we show the relic abundance for the heavy neutrino WIMP, $N_1$, for $v_{\chi^\prime}=3$~TeV and $4$~TeV. We should remark that the masses of $Z^{\prime}$ and $P_{1}$ depend only on the values of the VEVs, and will not change as we vary the several coupling constants in the model, while the $S_1$ mass contains an additional free coupling constant, $\lambda_1$. We then vary the $S_1$ mass instead of $\lambda_1$, and the range considered in this case is $400\ \mbox{GeV}\leq M_{S_1} \leq 4.5\ \mbox{TeV}$ for $v_{\chi^\prime}=3$~TeV and $600\ \mbox{GeV}\leq M_{S_1} \leq 6\ \mbox{TeV}$ for $v_{\chi^\prime}=4$~TeV.
Hence, the only relevant varying parameter (besides the neutrino mass) is $M_{S_1}$, which leads to a denser region in the abundance for large sterile neutrino masses.  
The region in accordance with WMAP7, $0.098 \leq \Omega h^2 \leq 0.122$, is shown between the red bars.
One can see that a change on $v_{\chi^\prime}$ is not going to affect appreciably the shape of the abundance, while it considerably change its quantitative aspect, diminishing the favored WMAP7 region for lower values of $v_{\chi^\prime}$. 

\begin{figure}[!htb]
\centering
\includegraphics[width=0.49\columnwidth]{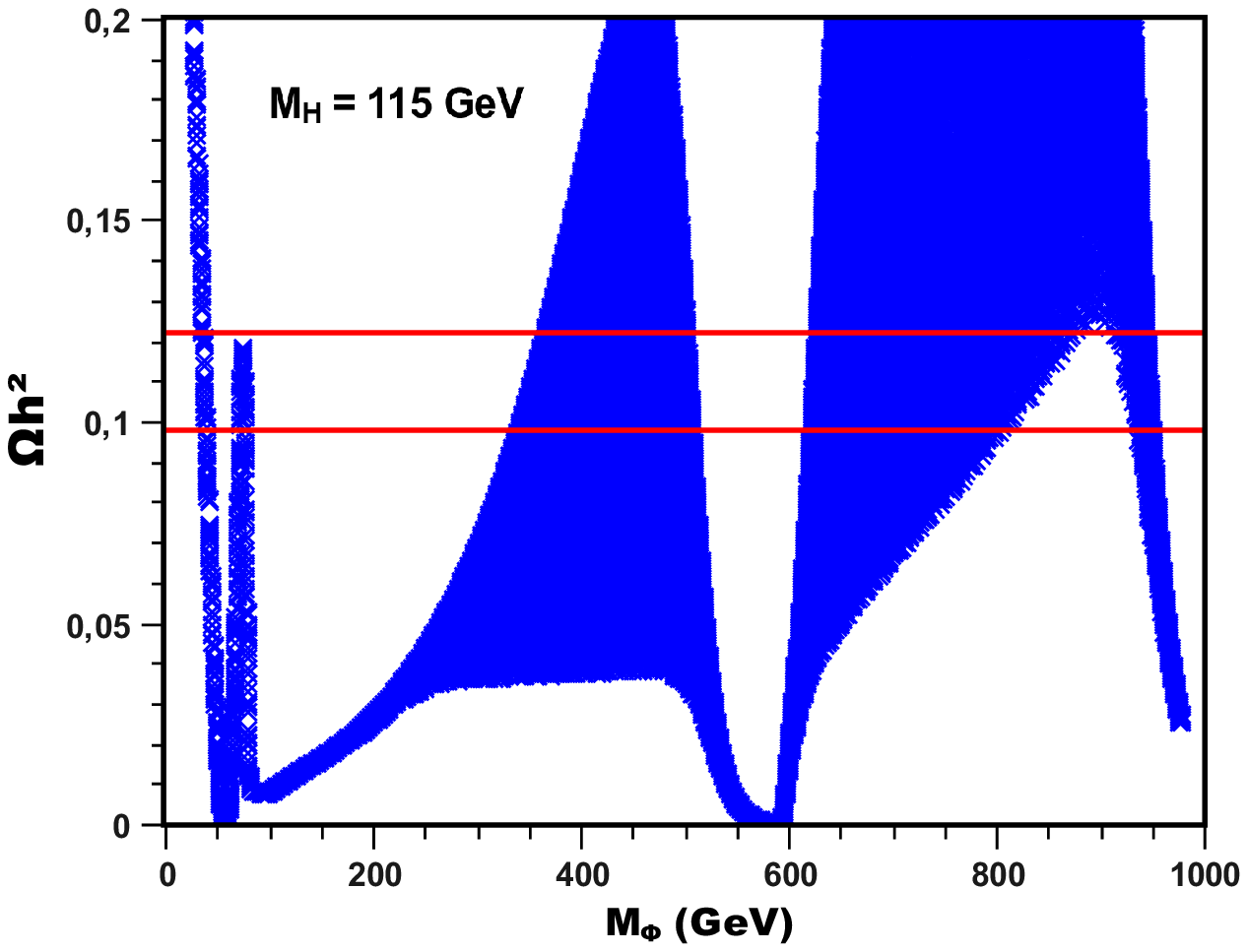}
\includegraphics[width=0.49\columnwidth]{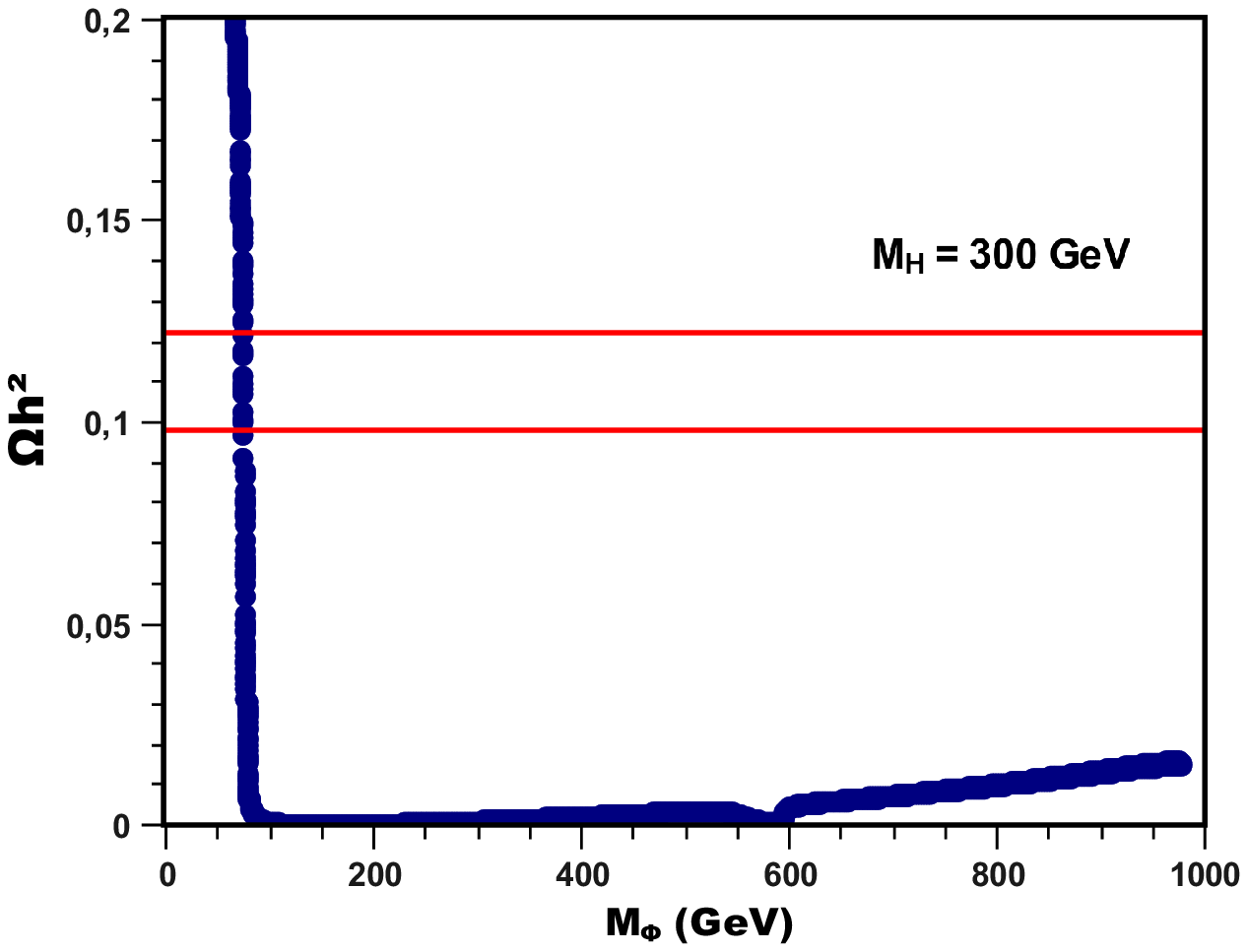}
\caption{The abundance of the scalar $\phi$ for two distinct values of the Higgs mass. The left panel is the abundance for $M_{H}=115\ \mbox{GeV}$ and the right one is for $M_{H}=300\ \mbox{GeV}$. We used $v_{\chi^\prime} = 3\ \mbox{TeV}$.}
\label{abundanciafi}
\end{figure}

As for the scalar $\phi$, using the same arguments used earlier for $N_{1}$, we observe that the only parameters which control its abundance are the $\phi$ mass and the masses of the Higgs and the scalar $S_{2}$. Let us remark that the $S_2$ mass depends on the same couplings as the Higgs mass and can be considered to be constant, since in the range of Higgs mass employed in this work, $115$~GeV to $300$~GeV, the $S_2$ mass change only about $5$~GeV. Hence, the abundance of $\phi$ is generally governed only by the Higgs mass. 
Nonetheless, the first process in Fig.~\ref{processofi} may be the most relevant when the produced quarks are heavy, and then we also vary the intermediate exotic quark mass parameter in the range $600$~GeV$\leq M_{q_i^\prime}\leq 2$~TeV.
In order to see the effect of varying the Higgs mass we show two plots in Fig.~\ref{abundanciafi} containing our results for the abundance with $M_H=115$~GeV and $M_H=300$~GeV.
Comparing the two panels we conclude that the abundance of $\phi$ is considerably modified by the Higgs mass, with a light Higgs boson offering a denser region on the parameter space. 

In summary, the model contains two interesting CDM candidates in two distinct regimes: one where $N_1$, a sterile neutrino, can account for the whole CDM and another where $\phi$, a scalar, is the CDM. Both can be stable (but not simultaneously, unless they are degenerate) thanks to a global $U(1)_G$ symmetry, under which only some of the new particles are charged,  implying that they are always produced in pairs, which resembles something like an R-parity, though it is related to the a continuous symmetry instead. Once observed that our candidates can account for the total CDM abundance, we need to check if they are in agreement with the last constraints from direct detection experiments and through this condition, we are going to assess a constraint on the symmetry breaking scale of the model.

We also want to check if there is some room to explain some of recent claims of a light CDM positive signal in CDMSII, which may be possible for the scalar $\phi$, whose mass can be made naturally small.

\subsection{Direct Detection}

After their decoupling, the WIMPs can cluster and form a local density of CDM surrounding us. Therefore the space at Earth location is supposed to be permeated by a flux of these particles characterized by a density and velocity distribution that depend on the details of the galactic halo model. If these WIMPs are allowed to interact with nuclei, through more fundamental interactions with quarks (for a good review see~\cite{reviewDM}), then it  is possible to directly detect them by measuring the recoil energy $(Q)$ deposited in the detector material, given by,
\begin{equation}
Q=2\frac{\mu_{r}^{2}v^{2}}{m_{N}},
\end{equation}
where $\mu_r=M_W m_N/(M_W + m_N)$ is the reduced WIMP-nucleus mass, $M_W$ the WIMP mass, $m_N$ the nucleus mass and $v$ is the minimal incoming velocity of an incident WIMP.

Measuring the energy deposited by the WIMP and making some assumptions about the halo model, we can infer the spin independent WIMP-nucleus cross section at zero momentum transfer, using the standard procedure described in \cite{micromegas,reviewDM,supersimetria2},

\be
\sigma_0 = \frac{4 \mu_r^2}{\pi}\left( Z f_p + (A-Z) f_n \right)^2\,.
\label{wimpnucleon}
\ee
where $Z$ is the atomic number, $A$ is the atomic mass and $f_{p}$ and $f_{n}$ are effective couplings with protons and neutrons, respectively, and depends of the particle physics input of a given model.

Since the DM experiments such as the Cryogenic DM Search (CDMS)~\cite{cdms} and the liquid noble gas XENON~\cite{xenon} contain nuclei with different atomic masses, 
its useful to define what we call the WIMP-nucleon cross section when $f_{p}\cong f_{n}$,
\be
\sigma_{p,n}^{SI} = \sigma_0 \frac{\mu_{p,n}^2}{\mu_r^2 A^2}\,.
\label{wimpnucleoncs}
\ee
where $\mu_{p,n}$ is the WIMP-proton/neutron reduced mass. 
The assumption $f_{p}\cong f_{n}$ is valid for most models, but there will be instances in our model where this fails to be true, as we will point later for the case of $N_1$.

These experiments have been trying to observe WIMP events, but in most of the cases no event have been detected and hence they were able to impose strong limits in the WIMP-nucleon cross section instead. Nevertheless, recently the CDMS collaboration has reported its results of the final data runs of the CDMSII and observed that two candidate events have survived after application of many discrimination procedures. The probability of observing two or more background events is $23\%$, which means that the two events neither provide a statistically significant evidence for CDM, nor can be rejected as  background. Many works have been done interpreting these two candidate events as WIMP signals in different frameworks~\cite{CDMSIImodels}. Here we will first investigate if the 3-3-1LHN CDM candidates satisfy the bounds imposed by these experiments and also
search for  an explanation to the events in excess observed by CDMSII.

The process which contribute to spin independent cross section of $N_{1}$ and $\phi$ are shown below, in Fig.~\ref{figN1} and \ref{figFi}.
\begin{figure}[htb]
\centering
\includegraphics[width=0.6\columnwidth]{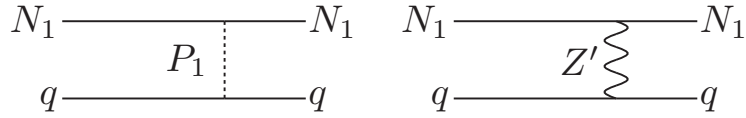}
\caption{Processes which contribute to the WIMP-nucleon cross section of $N_{1}$}
\label{figN1}
\end{figure}
\begin{figure}[htb]
\centering
\includegraphics[width=0.6\columnwidth]{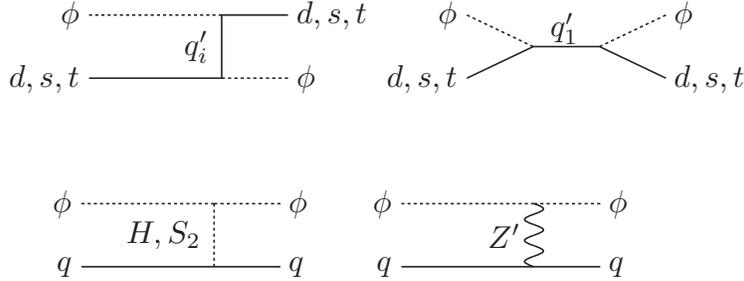}
\caption{Processes which contribute to the WIMP-nucleon cross section of $\phi$.}
\label{figFi}
\end{figure}
Well, after discussing a little bit about the direct detection method and showing the processes which contribute to the WIMP-nucleon cross section we are able to show and analyze the results for each candidate. 

\begin{figure}[!htb]
\centering
\includegraphics[width=0.6\columnwidth]{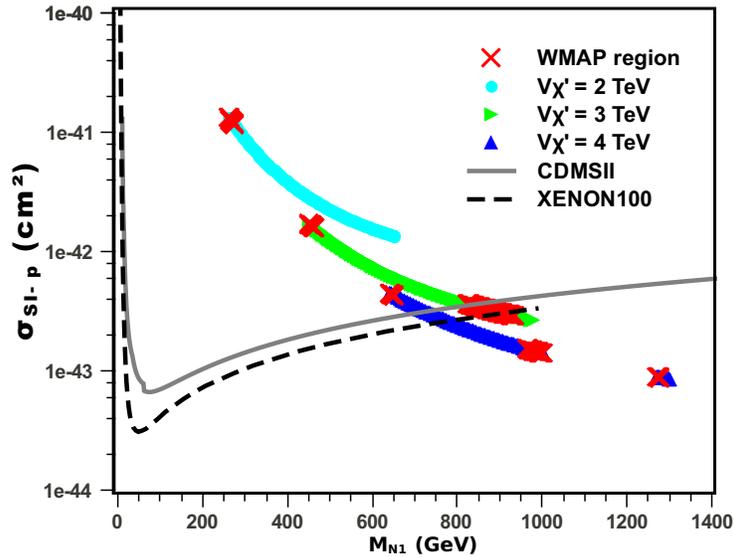}
\caption{The WIMP-proton cross section for $N_{1}$. From top to bottom, the curves represent the variation of $v_{\chi^\prime}$ in the range $2$~TeV$\leq v_{\chi^\prime} \leq 4\ \mbox{TeV}$. The data used in the exclusion curves were obtained using \cite{XENONtools}.}
\label{crosssectionN1}
\end{figure}

The scattering processes of $N_{1}$ with quarks are exhibited in the Fig.~\ref{figN1}. Using the fact that the vertices involving the gauge boson $Z^{\prime}$ possess only gauge couplings and that the scalar $P_1$ couples to $N_{1}$ proportionally to its mass, the only free parameters related to the WIMP-nucleon cross section of $N_{1}$ are its own mass and $v_{\chi^{\prime}}$. To see how our results are modified by the value of $v_{\chi^{\prime}}$ we evaluate the spin independent WIMP-nucleon cross section at zero momentum transfer limit given in Eq.~(\ref{wimpnucleoncs}), for $v_{\chi^{\prime}}$ varying from $2$~TeV to $4\ \mbox{TeV}$, which we present in the Fig.~\ref{crosssectionN1}.  Actually, in the case of $N_1$, the WIMP-nucleon coupling with protons is about one order of magnitude higher than the coupling with neutrons, and we choose to plot the WIMP-proton cross section since it is more strongly constrained than the neutron one in this case.

From the Fig.~\ref{crosssectionN1} we might realize that the heavy neutrino constitutes a nice CDM candidate obeying the most recent bounds from direct detection experiments if $v_{\chi^\prime}\geq 3$~TeV. The changing in the scale of symmetry breaking shows us that raising the values of $v_{\chi^\prime}$ we make the model safer if the experiment sensitivity grows. This is an interesting result for this model because the direct detection experiments are strongly constraining the breakdown of 3-3-1LHN symmetry to be above 3~TeV.
The gap in the results on this figure appears because it refers to the overabundant regime $(\Omega h^2 > 0.122)$ whose points were not included in the plot.

\begin{figure}[!htb]
\centering
\includegraphics[width=0.49\columnwidth]{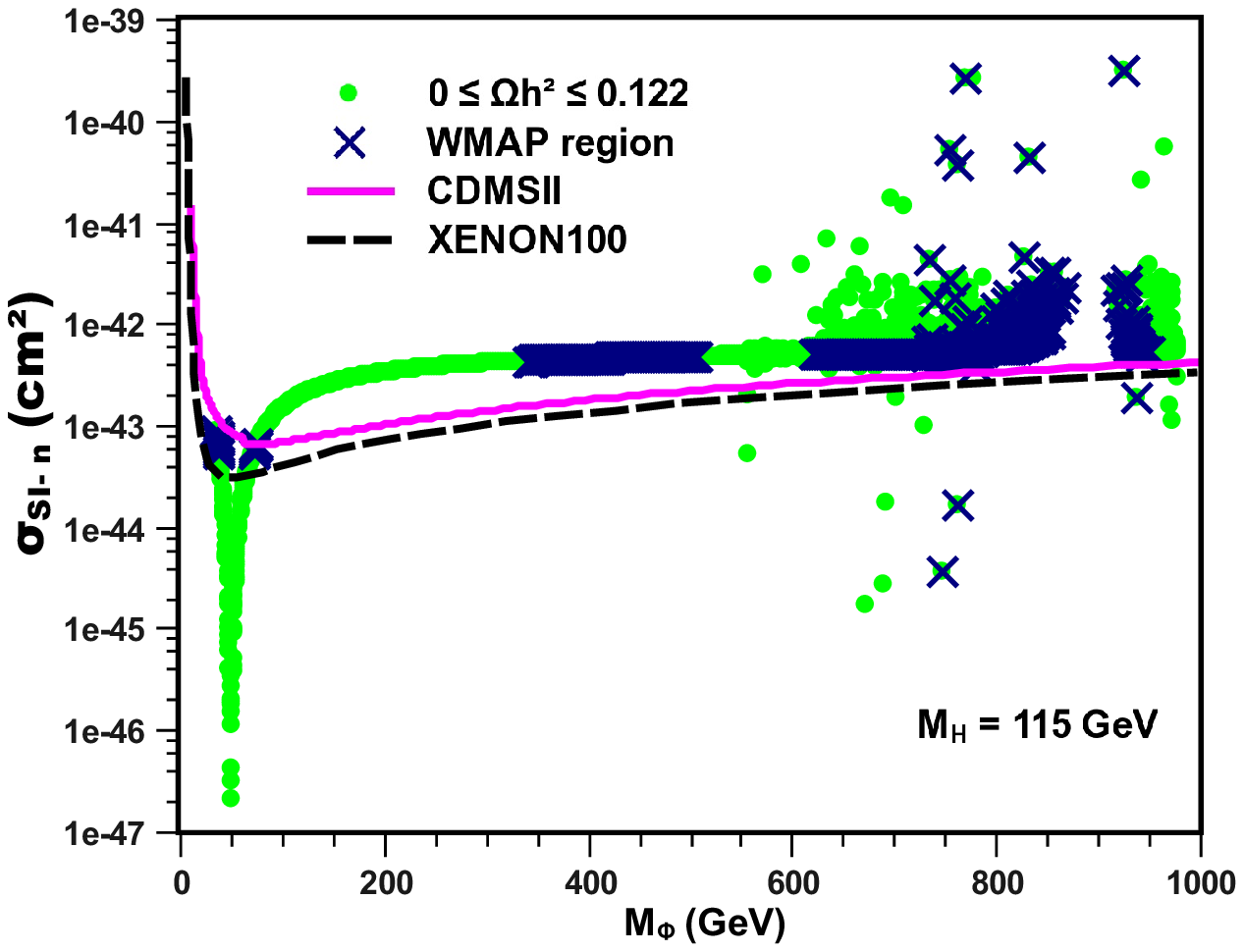}
\includegraphics[width=0.49\columnwidth]{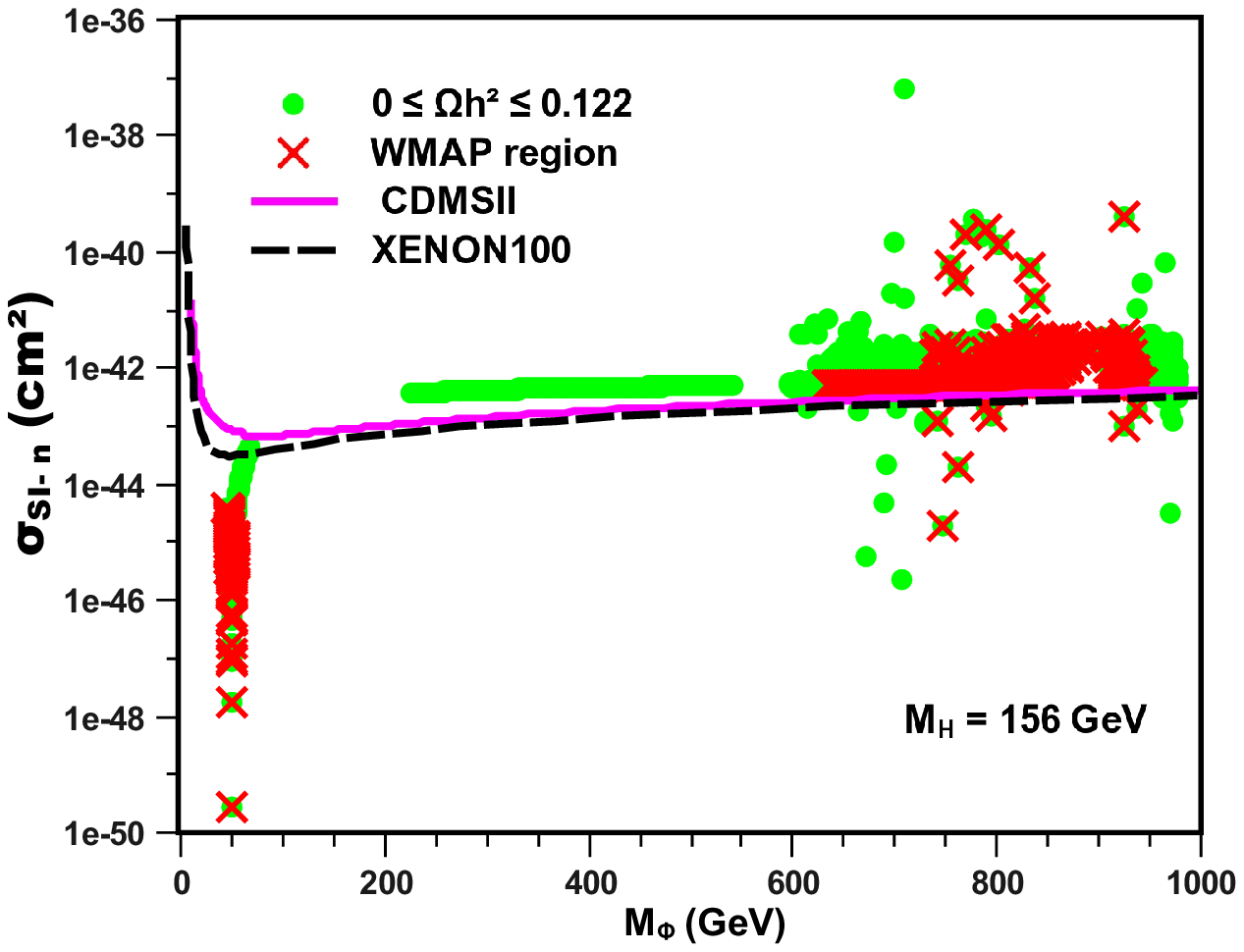}
\includegraphics[width=0.49\columnwidth]{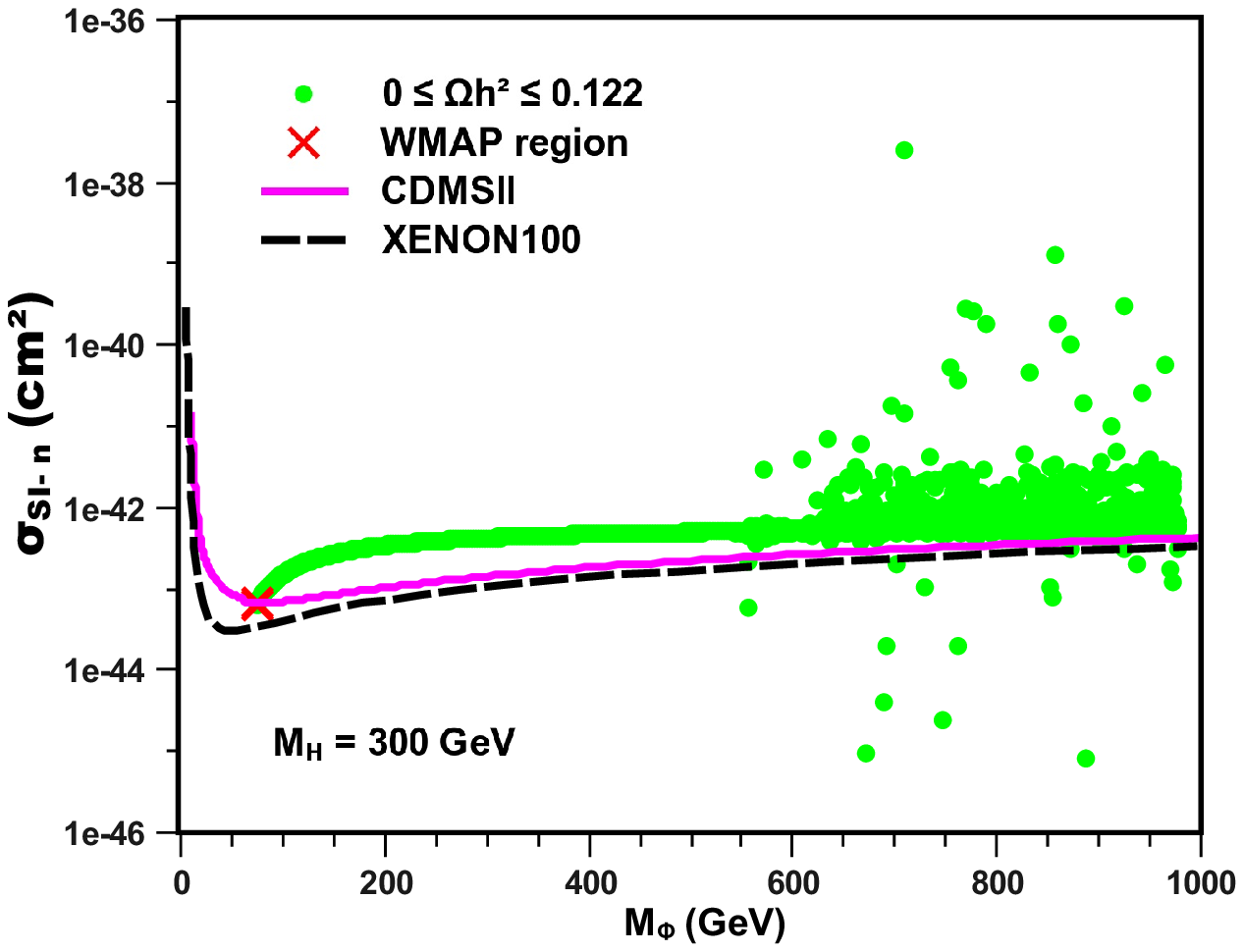} 
\includegraphics[width=0.49\columnwidth]{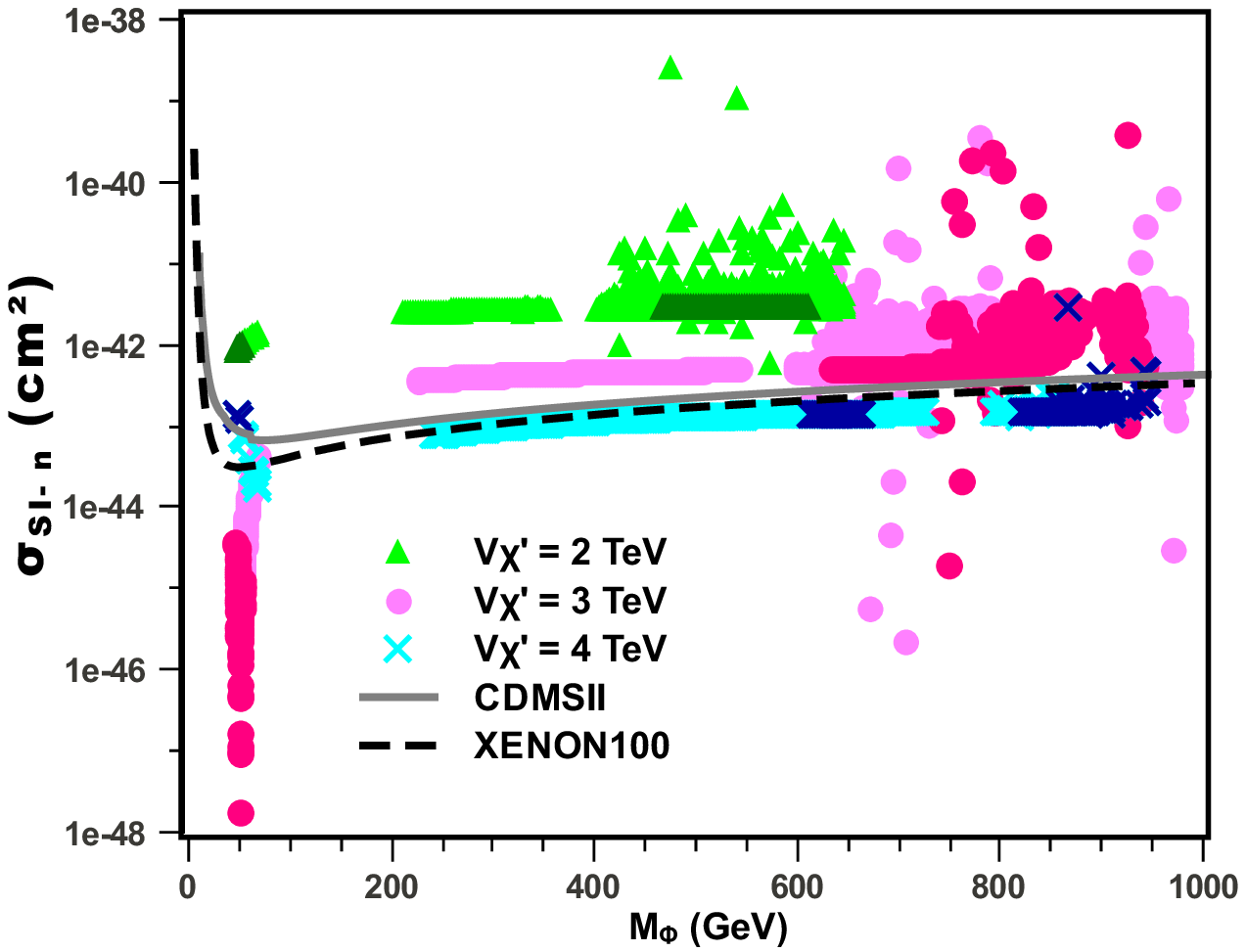}
\caption{The WIMP-nucleon cross section for $\phi$. The first panel is for $M_{H}= 115\ \mbox{GeV}$. The second one is for $M_{H}= 156\ \mbox{GeV}$. The third panel is for  $M_{H}= 300\ \mbox{GeV}$. We used $v_{\chi^\prime}= 3\ \mbox{TeV}$ in the first three plots while this parameter is varying and $M_H = 156$~GeV in the fourth plot, where the darker colors indicate those points in agreement with WMAP7, while lighter colors represent the region below the WMAP7 upper bound. The data used in the exclusion curves were obtained using \protect\cite{XENONtools}.}
\label{crosssectionfi}
\end{figure}

Finally for the scalar $\phi$, we can also calculate the WIMP-nucleon cross section taking into account the possible processes shown in the Fig.~\ref{figFi}. To understand how many free parameters are really important to the WIMP-nucleon cross section we provide some details in what follows. Since the exotic quark Yukawa couplings and scalar couplings are naturally of the order one, the cross section dependence on them can be translated into their masses, while for the $S_2$ scalar this reflects directly on $v_{\chi^\prime}$, since its mass is given by Eq.~(\ref{eigenvextorsRe}). 
Also, the gauge boson $Z^{\prime}$ contribution involves only gauge couplings. Therefore,  the only free parameters are the exotic quarks,  Higgs and $\phi$ masses, besides $v_{\chi^\prime}$. To precise how much the Higgs mass is important to our results we show the WIMP-nucleon cross section of $\phi$ for different Higgs masses in Fig.~\ref{crosssectionfi},

Comparing the first three plots in Fig.~\ref{crosssectionfi} we conclude that the WIMP-nucleon cross section of the scalar $\phi$ is very sensitive to changes in the Higgs mass and that the best parameter space in agreement with WMAP constraints is obtained for a light Higgs boson, in particular for a mass around $150\ \mbox{GeV}$. The spread points in each plot beyond $M_\phi\approx 500$~GeV are due to the changing in the masses of the exotic quarks in the range $636$~GeV$\leq M_{q_i^\prime}\leq 2$~TeV, which does not affect the cross section for lower WIMP masses. The cross section dependence on the $S_{2}$ mass, which is basically $v_{\chi^\prime}$, have an impact on the results as shown in the fourth panel of Fig.~\ref{crosssectionfi}, where we exhibit the WIMP-nucleon cross section behavior for different values of  $v_{\chi^\prime}$ and $M_H = 156$~GeV. In that plot the points in lighter colors represent a region of the parameter space corresponding to $\Omega h^2 \leq 0.122$, while the darker colors are the region in agreement with WMAP. We do this with the purpose of showing that direct detection bounds on the scalar $\phi$ seems to disfavor $v_{\chi^\prime}$ below $3$~TeV, as in the case of the sterile neutrino $N_1$. It is noticeable that most points for $v_{\chi^\prime} = 3$~TeV are ruled out for large WIMP masses, and a light $\phi$ is favored, while for $v_{\chi^\prime} = 4$~TeV the model is on the verge to be tested for the whole $\phi$ mass range. This reasonably high symmetry breaking scale is something to be taken into account when looking for 3-3-1LHN signals at LHC, a task we intend to perform soon.

\begin{figure}[!htb]
\centering
\includegraphics[width=0.6\columnwidth]{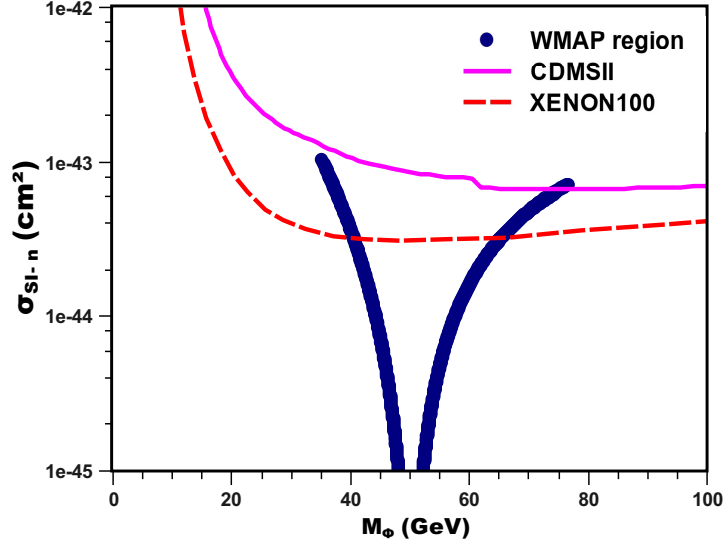}
\caption{The WIMP-nucleon cross section for low masses of $\phi$ with $115\ \mbox{GeV} \leq M_{H} \leq 300\ \mbox{GeV}$. We used $v_{\chi^\prime} = 3\ \mbox{TeV}$. The data used in the exclusion curves were obtained using \protect\cite{XENONtools}.}
\label{CDMSIIfi}
\end{figure}

In brief, we have checked that the two CDM candidates separately satisfy the exclusion limits from the most restrictive DM detection experiments. Now we will show that the model also provides an explanation to the two excess events observed by CDMSII Collaboration~\cite{cdms}. Since the scalar $\phi$ is the only candidate which can have low mass in agreement with these limits, it is the only one capable of representing those excess events. Computing again the WIMP-nucleon cross section only for low masses letting the Higgs mass free to vary from 115~GeV to 300~GeV, we obtain the behavior depicted in Fig.~\ref{CDMSIIfi},
We can observe that for the scalar $\phi$, there exists a region of parameter space with $M_\phi$ below $60\ \mbox{GeV}$ for which $10^{-43}$~cm$^{2} \leq \sigma_{SI}\leq 10^{-44} $~cm$^{2}$, that reproduces the two candidate events reported by CDMSII~\cite{cdms} and is not excluded by the recent bounds from XENON100. In addition to this, comparing the parameter space for distinct values of the Higgs mass we conclude that if these events are really a WIMP signal due to the scalar $\phi$, it prefers a Higgs boson with $M_H\approx 150$~GeV (see Fig.~\ref{crosssectionfi}) and in this way the solution of the DM problem brings to us some hints on Higgs search. It is important to say that the results shown above were obtained with only two free parameters, which are the scalar potential coupling constant, $\lambda_{7}$, and the Higgs mass. 

Finally, we should mention that this work has not only enlarged the possibilities of candidates in the 3-3-1 model, as compared to Ref.~\cite{JCAP}, but also a deeper analysis was carried out considering a wider range of parameter space. This was achieved by the introduction of a heavy neutrino into the spectrum allowing for a new global $U(1)_G$ symmetry that would no be possible in the 3-3-1 model with right-handed neutrinos~\cite{JCAP}. This new symmetry was crucial to establish that the lightest new particles charged under $U(1)_G$ are stable.

\section{Conclusions}
\label{sec4}

We have studied a 3-3-1 model with heavy sterile neutrinos and observed that the model accommodates a new extra global $U(1)_G$ symmetry, that makes possible the identification of three CDM candidates in its mass spectrum. One of them, a non-hermitian vector boson $U^0$ was not considered in our analysis because it does not provide enough amount of CDM. The remaining two are a neutral scalar, $\phi$, studied before in another version of the 3-3-1 model~\cite{JCAP} in a very restricted scenario (and stable thanks to a lepton number symmetry), and the lightest of the heavy neutrinos, $N_1$. We have shown that the scalar $\phi$ and the sterile neutrino $N_1$ can account for the total CDM in agreement with WMAP7 data. 
We then computed the scattering cross section of our WIMPs with nucleons, in order to comply with recent direct detection experiments, CDMSII and XENON100, and concluded that there is a large range of the parameter space that obeys their exclusion limits. For the scenario where $\phi$ is the WIMP DM we also concluded that a Higgs mass of about 150~GeV is favored by these limits. 

Besides, an interesting outcome has emerged from our analysis concerning the direct detection of our CDM WIMPs, $N_1$ and $\phi$, which is the fact the characteristic symmetry breaking scale of the 3-3-1RHN model, $v_{\chi^\prime}$, has to be larger than about~3~TeV so as to evade the current exclusion limits from CDMSII and XENON100. This is an important feature to be considered in testing this model at LHC (a work to be developed elsewhere).

Finally, we saw that the scalar $\phi$ might reproduce the two excess events reported by CDMSII in the Fig.~\ref{CDMSIIfi}.  Our results imply that the model can either satisfy the exclusion limits and/or explain the positive signal observed by CDMSII, pointing to a Higgs mass below $300$~GeV with strong bounds on the 3-3-1RHN symmetry breakdown scale.

\section*{Acknowledgments}

We feel deeply indebted to A. Semenov for valuable information about LanHEP and CalcHEP and A. Pukhov for his prompt help concerning micrOMEGAs package. This research was supported by the Conselho Nacional de Desenvolvimento Cient\'{\i}fico e Tecnol\'ogico (CNPq) (CASP and PSRS) and Coordena\c{c}\~ao de aperfei\c{c}oamento de Pessoal de N\'{\i}vel Superior (CAPES) (FSQ). 

\appendix
\section*{Appendix A}

In the tables below we present all the triple interactions and couplings involving the $\mathbf{G}$-fields in the 3-3-1LHN model, relevant to determine the stability of our CDM candidates. Here we define, $e$ as the electric charge and,

$R_1=\sqrt{1+\frac{v^2}{v_{\chi^\prime}^2}}$,\ $R_2=\sqrt{2+\frac{v^2}{v_{\chi^\prime}^2}}$,\ $g_W=1-2s^2_W$, $\alpha_1=3-4s^2_W$

${\bf t_N}=\frac{\sqrt{3}s_W}{\sqrt{3-4s^2_W}}$,\ ${\bf q}=\frac{9}{3-4s^2_W}$,\ ${\bf s}=\frac{9c^2_W}{3-4s^2_W}$,\ ${\bf p}=9\frac{(1-2s^2_W)}{3-4s^2_W}$.

For simplicity, no hermitian conjugate interaction is included in these tables and interactions that are already included in one table are not present in the others.

\begin{tabular}{|l|l|}
\hline
$\phi$ interactions & Couplings \\
\hline  \hline
 $\phi\ \phi^{\star}\ H$ & $-\frac{\sqrt{2}v}{2R^2_1} \left( 2\lambda_2 + \lambda_6 + \lambda_7 + \frac{1}{2} + \frac{v^2}{v_{\chi^\prime}^2}(\lambda_4+\lambda_5 + \lambda_7 ) \right)$\\ \hline
 $\phi\ H\ U^{0}_\mu$ & $-\frac{g\sqrt{2}}{4R_1} (p_1 - p_2)_\mu $\\ \hline
 $ \phi\ \overline{N}_a\ \nu_a $ & $-\frac{ g^{\prime}_{aa} v}{2V R_1} (1- \gamma_5)$ \\ \hline
 $\phi\ P_1\ U^{0}_\mu$ & $ \frac{ig}{2R_1 R_2} (p_1-p_2)_\mu $ \\ \hline
 $\phi\ \overline{u^{\prime}}_3\ t$ & $-\frac{1}{2 R_1}\left(\frac{v}{v_{\chi^\prime}}f_{33}(1-\gamma_5) + \frac{ m_t\sqrt{2}}{v} (1+\gamma_5)\right)$ \\ \hline
 $\phi\  \bar{s}\ d^{\prime}_2$ & $-\frac{1}{2 R_1}\left(\frac{v}{v_{\chi^\prime}}f_{22}(1+\gamma_5) + \frac{ m_s\sqrt{2}}{v} (1-\gamma_5)\right)$ \\ \hline
$\phi\ \bar{d}\ d^{\prime}_1 $ & $-\frac{1}{2 R_1}\left(\frac{v}{v_{\chi^\prime}}f_{11}(1+\gamma_5) + \frac{ m_d\sqrt{2}}{v} (1-\gamma_5)\right)$ \\ \hline
 $ \phi\ \phi^{\star}\ S_2$ & $-\frac{\sqrt{2v}}{2R_1}\left( 2\lambda_2 - \lambda_6 +\lambda_7 -1/2 + \frac{v^2}{v_{\chi^\prime}^2}(\lambda_4-\lambda_5+\lambda_7)\right)$ \\ \hline
 $ \phi\ S_2\ U^{0}_\mu$ & $-\frac{\sqrt{2}g}{4R_1}(p_1 - p_3)_\mu $ \\ \hline
 $ \phi\ \phi^{\star}\ S_1$ & $-\frac{v}{R_1}\left( \frac{v}{v_{\chi^\prime}}(2\lambda_1+\lambda_7)+\frac{v_{\chi^\prime}}{v}(\lambda_4+\lambda_7)\right)$ \\ \hline
 $\phi\ S_1\ U^{0}_\mu$ & $-\frac{gv}{v_{\chi^\prime} R_1} p_{2 \mu}$ \\ \hline
 $\phi\ V^{-}_\mu\ W^+_\nu $ & $ \frac{\sqrt{2} g^2 v}{2R_1} g_{\mu \nu} $ \\ \hline
 $\phi\ h^-_2\ V^{-}_\mu $ & -$\frac{g}{2R_1}(p_1 - p_2)_\mu $ \\ \hline
 $\phi\ Z^\mu\ U^{0}_\nu $ & $\frac{g^2 v \sqrt{ {\bf q} }}{2 R_1\sqrt{ {\bf s} }}g_{\mu \nu} $ \\ \hline
 $\phi\ Z^{\prime}_\mu\ U^{0}_\nu $ & $\frac{g^2 v}{6 R_1\sqrt{ {\bf s}  }}({\bf p}-2{\bf s})g_{\mu \nu} $ \\ \hline
 $\phi\ \phi^{\star}\ Z^{\prime \mu} $ & $\frac{g \sqrt{ {\bf s} }}{3R^2_1 } (p_1 - p_2)_{\mu}$ \\ \hline
 $\phi\ h_1^-\ h_2^+$ & $-\frac{v}{2 R^2_1}\left(\frac{v^2}{v_{\chi^\prime}^2}(\lambda_7+\lambda_8)+(\lambda_7+\lambda_8+2\lambda_9-1)\right)$ \\ \hline
 \end{tabular}
\vspace{1cm}

\begin{tabular}{|l|l|}
\hline
$N_1$ interactions & Couplings \\
\hline  \hline
 $\overline{e}\ N_1\  V_\mu^-$ & $-\frac{g\sqrt{2}}{4}\gamma_\mu(1-\gamma_5) $\\ \hline 
 $\overline{e}\ N_1\ h_1^-$ & $-\frac{1}{2R_1}\left( \frac{m_e\sqrt{2}}{v}(1-\gamma_5) + \frac{v}{v_{\chi^\prime}}g^{\prime}_{11}(1+\gamma_5)\right)$\\ \hline
 $\overline{N}_1\ N_1\ P_1$ & $- \frac{ig^{\prime}_{11}v\sqrt{2}}{2v_{\chi^\prime} R_2}\gamma_5 $\\ \hline
 $\overline{N}_1\  N_1\ S_1 $ & $-\frac{g^{\prime}_{11}\sqrt{2}}{2}$ \\ \hline
 $\overline{N}_1\ N_1\ Z^{\prime}_\mu$ & $ \frac{g}{6\sqrt{ {\bf s} }}(3 + {\bf t^2_N} ) \gamma_\mu (1-\gamma_5) $ \\ \hline
 $\overline{\nu}_e\ N_1\ U^{0}_\mu$ & $-\frac{g\sqrt{2}}{4}\gamma_\mu (1-\gamma_5)$ \\ \hline
 $\overline{\nu}_e\ N_1\ \phi^{\star}$ & $-\frac{g^{\prime}_{11} v}{2v_{\chi^\prime}R_1}$ \\ \hline
 \end{tabular}
\newpage

\begin{tabular}{|l|l|}
\hline
$U^0$  interactions & Couplings \\
\hline  \hline
 $U^0_\mu\ U^{0 \star}_\nu\ H$ & $-\frac{\sqrt{2}g^2v}{4}g_{\mu \nu} $\\ \hline 
 $U^{0 \star}_\mu\ \overline{N_a}\ \nu_a$ & $-\frac{\sqrt{2}g}{4} \gamma_\mu(1-\gamma_5)$ \\ \hline
 $U^0_\mu\ \overline{d^{\prime}_1}\ d$ & $\frac{\sqrt{2}g}{4} \gamma_\mu(1-\gamma_5) $ \\ \hline
 $U^0_\mu\ \overline{d^{\prime}_2}\ s$ & $\frac{\sqrt{2}g}{4} \gamma_\mu(1-\gamma_5)$ \\ \hline
 $U^{0 \star}_\mu\ \overline{u^{\prime}_3}\ t$ & $-\frac{\sqrt{2}g}{4} \gamma_\mu(1-\gamma_5)$ \\ \hline
 $U^0_\mu\ U^{0 \star}_\nu\ S_2$ &   $\frac{g^2\sqrt{2}v}{4}g_{\mu \nu}$ \\ \hline
 $U^0_\mu\ U^{0 \star}_\nu\ S_1$ &   $\frac{g^2v_{\chi^\prime}}{2}g_{\mu \nu}$ \\ \hline
 $U^0_\rho\ V^+_\mu\ W^-_\nu$ &¨$-\frac{\sqrt{2}g}{2}\left( p_{1\nu} g_{\mu \rho} - p_{1\mu} g_{\nu \rho} - p_{2\nu} g_{\mu \rho} + p_{2\rho} g_{\mu \nu} + p_{3\mu} g_{\nu \rho} - p_{3\rho} g_{\mu \nu} \right)$ \\ \hline
 $U^0_\mu\ V^+_\nu\ h_2^-$ & $\frac{g^2 v}{2}g_{\mu \nu}$ \\ \hline
 $U^0_\mu\ h_1^+\ W^-_\nu$ & $\frac{g^2 \sqrt{2}}{2R1}g_{\mu \nu}$ \\ \hline
 $U^0_\rho\ U^{0 \star}_\mu\ Z_\nu $ & $\frac{g}{2C_W}\left(p_{1 \mu} g_{\nu \rho} - p_{1\nu} g_{\rho \mu} - p_{2 \rho}g_{\nu \mu} + p_{2 \nu}g_{\rho \mu} +p_{3 \rho}g_{\nu \mu} - p_{3 \mu} g_{\nu \rho} \right)$ \\ \hline
 $U^0_\rho\ U^{0 \star}_\mu\ Z^{\prime}_\nu $ & $\frac{\alpha_1 g}{2CW}\left(p_{1 \mu} g_{\nu \rho} - p_{1\nu} g_{\mu \rho} - p_{2 \rho}g_{\mu \nu} + p_{2 \nu}g_{\rho \mu} + p_{3 \rho}g_{\nu \mu} - p_{3 \mu} g_{\nu \rho} \right)$ \\ \hline
$U^{0 \star}_\mu\ h_1^-\ h_2^+$ & $-\frac{g}{2R1}(p_3-p_2)_\mu$ \\ \hline
\end{tabular}
\vspace{1cm}

\begin{tabular}{|l|l|}
\hline
$V^\pm$ interactions & Couplings \\
\hline  \hline
 $V^+_\mu\ V^-_\nu\ A_\rho$ & $-e\left(p_{1 \rho}g_{\mu \nu} - p_{1\nu}g_{\rho \mu}
  +p_{2\mu}g_{\rho \nu} - p_{2\rho}g_{\mu \nu} + p_{3 \nu}g_{\rho \mu} - p_{3\mu}g_{\rho \nu}\right) $\\ \hline 
 $V^+_\mu\ \bar{u}\ d^{\prime}_1 $ & $-\frac{g\sqrt{2}}{4}\gamma_\mu (1-\gamma_5)$ \\ \hline
 $V^+_\mu\ \bar{c}\ d^{\prime}_2 $ & $-\frac{g\sqrt{2}}{4}\gamma_\mu (1-\gamma_5)$ \\ \hline
 $V^+_\mu\ \overline{u^{\prime}_3}\ b$ & $-\frac{g\sqrt{2}}{4}\gamma_\mu (1-\gamma_5)$ \\ \hline
$V^+_\mu\ V^-_\nu\ H$ & $\frac{g^2v\sqrt{2}}{4}g_{\mu \nu}$ \\ \hline
$V^{+}_\mu\ h_1^{-}\ H$ & $-\frac{g\sqrt{2}}{4R_1}(p_2-p_3)_\mu$ \\ \hline
$V^{+}_\mu\ h_1^{-}\ P_1$ & $\frac{ig}{2R_1 R_2}(p_2-p_3)_\mu$ \\ \hline
$V^{+}_\mu\ V^{-}_\nu\ S_2$ & $-\frac{g^2 v \sqrt{2}}{4}g_{\mu \nu}$ \\ \hline
$V^{+}_\mu\ h_1^{-}\ S_2$ & $\frac{g\sqrt{2}}{4R_1}(p_2-p_3)_\mu$ \\ \hline
 $V^+_\mu\ V^-_\nu\ S_1$ & $\frac{g^2v_{\chi^\prime}}{2}g_{\mu \nu}$ \\ \hline
$V^+_\mu\ h_1^-\ S_1$ & $-\frac{gv}{v_{\chi^\prime}\ 2R_1}p_{3\mu}$ \\ \hline
$V^+_\mu\ V^-_\nu\ Z_\rho$ & $-\frac{g}{2C_W}(p_{2\mu}g_{\nu \rho}g_W - p_{2\rho}g_{\mu \nu}g_W - p_{1\nu}g_{\mu \rho}g_W + p_{1 \rho}g_{\mu \nu}g_W + p_{3\nu}g_{\mu \rho}g_W - p_{3 \mu}g_{\nu \rho}g_W)$ \\ \hline
$V^+_\mu\ V^-_\nu\ Z^{\prime}_\rho$ & $\frac{\alpha_{1} g}{2C_W}(p_{2\mu}g_{\nu \rho}- p_{2\rho}g_{\mu \nu} - p_{1\nu}g_{\mu \rho} + p_{1 \rho}g_{\mu \nu} + p_{3\nu}g_{\mu \rho}- p_{3 \mu} g_{\nu \rho})$ \\ \hline
$V^+_\mu\ Z_\nu\ h_1^-$ & $-\frac{g^2v}{4R_1 \sqrt{{\bf qs }}}({\bf p}  + {\bf q} +6 {\bf t}^2_{{\bf N}})g_{\mu \nu}$ \\ \hline
$V^+_\mu\ Z^{\prime}_\nu\ h_1^-$ & $\frac{g^2v}{12R_1\sqrt{ {\bf s} }}(-2 {\bf s} + {\bf q} - {\bf p})g_{\mu \nu}$ \\ \hline 
\end{tabular} \\

\begin{tabular}{|l|l|}
\hline
$h_1^\pm$ interactions & Couplings \\
\hline  \hline
 $h_1^-\ h_1^+\ A_\mu$ & $-\frac{e}{R^2_1}( p_2 - p_1)_\mu$\\ \hline 
 $h_1^-\ h_1^+\ H$ & $-\frac{\sqrt{2}v}{2R^2_1}\left(2\lambda_3+\lambda_6+\lambda_8 +1/2+\frac{v^2}{v_{\chi^\prime}^2}(\lambda_4+\lambda_5+\lambda_8\right)$\\ \hline
 $h_1^-\ P_1\ V^+_\mu$ & $\frac{ig}{2R_1R_2}(p_1-p_2)_\mu$ \\ \hline
 $h_1^-\ \overline{d^{\prime}_1}\ u$ & $\frac{1}{2R_1}\left(\frac{v}{v_{\chi^\prime}} f_{11} (1-\gamma_5) +\frac{m_u\sqrt{2}}{v}(1+\gamma_5)\right)$ \\ \hline
 $h_1^-\ \overline{d^{\prime}_2}\ c$ & $\frac{1}{2R_1}\left(\frac{v}{v_{\chi^\prime}} f_{22} (1-\gamma_5) +\frac{m_c\sqrt{2}}{v}(1+\gamma_5)\right)$ \\ \hline
 $h_1^-\ \bar{b}\ u^{\prime}_3 $ & $-\frac{1}{2R_1} \left(\frac{v}{v_{\chi^\prime}}f_{33}(1+\gamma_5)+\frac{m_b\sqrt{2}}{v}(1-\gamma_5)\right)$\\ \hline
 $h_1^-\ h_1^+\ S_2$ & $\frac{\sqrt{2}v}{2R^2_1}\left(2\lambda_3-\lambda_6+\lambda_8-1/2+\frac{v^2}{v_{\chi^\prime}^2}(\lambda_5+\lambda_8-\lambda_4)\right)$  \\ \hline
 $h_1^-\ h_1^+\ S_1$ & $\frac{-v}{R^2_1}\left( \frac{v}{v_{\chi^\prime}}(2\lambda_1+\lambda_8) + \frac{v_{\chi^\prime}}{v}(\lambda_5+\lambda_8)\right)$  \\ \hline
 $h_1^-\ h_1^+\ Z_\mu$ & $\frac{3g\ {\bf t}^2_{\bf N}}{R^2_1 \sqrt{ {\bf q\ s}   }} (p_2- p_1)_\mu$ \\ \hline
 $h_1^-\ h_1^+\ Z^{\prime}_\mu$ & $\frac{g\ {\bf p}}{3 R^2_1 \sqrt{ {\bf s} }} (p_2- p_1)_\mu$ \\ \hline
 \end{tabular}
\vspace{1cm}

\begin{tabular}{|l|l|}
\hline
$q^{\prime}_a$ interactions & Couplings \\
\hline  \hline
 $\overline{q^{\prime}_a}\ q^{\prime}_a\ A_\mu$ & $-Q_{q_a}\ e\gamma_\mu $\\ \hline 
$\overline{q^{\prime}_a}\ q^{\prime}_a\ P_1$ & $\frac{iv\sqrt{2}f_{aa}}{2v_{\chi^\prime}\ R_2}\gamma_5$ \\ \hline
$\overline{q^{\prime}_a}\ q^{\prime}_a\ S_1$ & $-\frac{\sqrt{2}f_{aa}}{2}$ \\ \hline
$\overline{q^{\prime}_a}\ q^{\prime}_a\ Z_\mu$ & $-\frac{g\ {\bf t}^2_{\bf N}}{2 \sqrt{ {\bf q\ s} }}\gamma_\mu$ \\ \hline
$\overline{q^{\prime}_a}\ q^{\prime}_a\ Z^{\prime}_\mu$ & $-\frac{g}{6 \sqrt{ {\bf s} }}(3\gamma_\mu (1-\gamma_5)-{\bf t}^2_{\bf N}\gamma_\mu (1+\gamma_5)) $ \\ \hline
\end{tabular}
\vspace{1cm}

{\section*{References}}

\end{document}